# A New Algebraic Approach to Perturbation Theory

**B. Gönül, N. Çelik and E. Olğar**

Department of Engineering Physics, University of Gaziantep, 27310, Gaziantep-Türkiye

**Abstract**

An algebraic non-perturbative approach is proposed for the analytical treatment of Schrödinger equations with a potential that can be expressed in terms of an exactly solvable piece with an additional potential. Avoiding disadvantages of standard approaches, new handy recursion formulae with the same simple form both for ground and excited states have been obtained. As an illustration the procedure, well adapted to the use of computer algebra, is successfully applied to quartic anharmonic oscillators by means of very simple algebraic manipulations. The trend of the exact values of the energies is rather well reproduced for a large range of values of the coupling constant $(g = 0.001 - 10000)$

Keywords: Perturbation theory, anharmonic oscillator

PACS No: 03.65.Fd.

## 1. INTRODUCTION

The main task in application of the quantum mechanics is to solve Schrödinger equations with different potentials. Unfortunately, realistic physical problems can practically never be solved exactly. Then one has to resort to some approximations. Most widely used among them is the perturbation theory. However, the explicit calculation with the Rayleigh-Schrödinger perturbation theory, described in most quantum mechanics textbooks, runs into the difficulty of the summation over all intermediate unperturbed eigenstates. To avoid this difficulty, alternative perturbation procedures have been proposed, notably the logaritmic perturbation theory (LPT) [1-4] and the Dalgarno-Lewis technique [5-8]. The virtue of LPT is its avoidance of the cumbersome summation over states for second- and higher-order corrections in Rayleigh-Schrödinger perturbation theory. Unfortunately, it has problems of its own in calculating corrections to excited states, owing to presence of nodes in the wavefunctions. Various schemes have been proposed to circumvent the resulting singularities [4, 9, 10].

Considering such drawbacks of the available treatments and gaining confidence from the success of the recent works [11], this Letter presents an alternative approach to perturbation theory in one-dimensional non-relativistic quantum mechanics, which yields simple but closed perturbation theory formulae leading to the Riccati equation from which one can actually obtain all the perturbation corrections to both energy level shifts and wavefunctions for all state. These quantities can be calculated to any given accuracy since the generation of successive corrections in the present perturbative expansions only requires the solution of simple algebraic solutions. The model applicable in the same form to both the ground state and excited bound states without involving tedious calculations which appear in the available perturbation theories. In particular it is noted that the procedure introduced here does not involve either tedious explicit factoring out of the zeros of [1,2] or introduction of ghost states [4] as were the cases encountered for applying LPT to excited states. In the application of the present method to the $n$th excited state, one requires only knowledge of the unperturbed state eigenfunction but no knowledge of the other eigenvalues or eigenfunctions is necessary.

As an illustration, the present scheme is applied to quartic anharmonic oscillator since there has been a great deal of interest in the analytical and numerical investigation of the one-dimensional anharmonic oscillator. They are of interest because of their importance in molecular vibrations [12] as well as in solid state physics [13] and quantum field theories [14]. Since anharmonic oscillators model intrinsic anharmonic effects of the real world, they continue to play a crucially important role in contemporary physics. On the other hand, the anharmonic oscillators with quartic potentials can serve as a testing ground for the various methods based on perturbative and non-perturbative approaches. In other words, interest in such a model stems mainly from the fact that, if one considers the anharmonicity $gx^4$ as a perturbing operator, then the Rayleigh-Schrödinger perturbation expansion for the eigenvalues diverges [15] for every value of $g$. Consequently, several methods have been used to calculate the quartic anharmonic oscillator eigenvalues and eigenfunctions. Without being exhaustive, we may recall variational methods [16], WKB methods [17], Hill determinat [18-19], Riccati [20] or Riccati-Hill determinat methods [21], perturbative treatment prescriptions using summability techniques such as the Stieljes, Pade and Borrell methods [14,22]. Let us also mention the hypervirial perturbation method of Fernandez and Castro [23], which can be viewed as a generalization of the Killingbeck method [24], and other alternative treatments [25], together with those involving a group-theoretical approach [26], the multiple scale technique [27], and supersymmetric methods [28]. Afterall, it appears challenging to test our formalism in avoiding the failure of the other perturbation series for the treatment of the quartic anharmonic oscillator.

The layout of the paper is as follows. In the next section we summarize the main ideas of our approach. The application of the present model to quartic anharmonic oscillators leading to simple recursion relations for the calculations at each succesive perturbation order and the results obtained are shown in section 3. The paper ends with a brief summary and concluding remarks.

## 2. THE MODEL

We first start with a brief introduction of the present formalism. Throughout the paper the unit system $\hbar = 2m = 1$ is chosen. In general, the goal in the supersymmetric quantum theory [29] is to solve the Riccati equation,

$$W^2(r) - W'(r) = V(r) - E_0, \qquad (1)$$

where $V(r)$ is the potential of interest and $E_0$ is the corresponding ground state energy. If we find $W(r)$, the so called superpotential, we have of course found the ground state wave function via,

$$\psi_0(r) = N \exp\left[-\int^r W(z)dz\right], \qquad (2)$$

where $N$ is the normalization constant. If $V(r)$ is a shape invariant potential, we can in fact obtain the entire spectrum of bound state energies and wave functions via ladder operators.

Keeping in mind this point, now suppose that we are interested in a potential for which we do not know $W(r)$ exactly. More specifically, we assume that $V(r)$ differs by a small amount from a potential $V_0(r)$ plus angular momentum barrier if any, for which one solves the Riccati equation explicitly. For the consideration of spherically symmetric potentials, the corresponding Schrödinger equation for the radial wave function has the form

$$\frac{\psi_n''(r)}{\psi_n(r)} = [V(r) - E_n] \quad , \quad V(r) = \left[V_0(r) + \frac{\ell(\ell+1)}{r^2}\right] + \Delta V(r), \qquad (3)$$

where $\Delta V$ is a perturbing potential. Let us write the wave function $\psi_n$ as

$$\psi_n(r) = \chi_n(r)\phi_n(r) \ , \tag{4}$$

in which $\chi_n$ is the known normalized eigenfunction of the unperturbed Schrödinger equation whereas $\phi_n$ is a moderating function corresponding to the perturbing potential. Substituting (4) into (3) yields

$$\left(\frac{\chi_n''}{\chi_n} + \frac{\phi_n''}{\phi_n} + 2\frac{\chi_n'}{\chi_n}\frac{\phi_n'}{\phi_n}\right) = V - E_n \ . \tag{5}$$

Instead of setting the functions $\chi_n$ and $\phi_n$, we will set their logarithmic derivatives using the spirit of Eqs. (1-2) and the standard approach of LPT ;

$$W_n = -\frac{\chi_n'}{\chi_n} \ , \quad \Delta W_n = -\frac{\phi_n'}{\phi_n} \tag{6}$$

which leads to

$$\frac{\chi_n''}{\chi_n} = W_n^2 - W_n' = \left[V_0(r) + \frac{\ell(\ell+1)}{r^2}\right] - \varepsilon_n \ , \tag{7}$$

where $\varepsilon_n$ is the eigenvalue of the exactly solvable unperturbed potential, and

$$\left(\frac{\phi_n''}{\phi_n} + 2\frac{\chi_n'}{\chi_n}\frac{\phi_n'}{\phi_n}\right) = \Delta W_n^2 - \Delta W_n' + 2W_n \Delta W_n = \Delta V(r) - \Delta \varepsilon_n \ , \tag{8}$$

in which $\Delta \varepsilon_n$ is the eigenvalue for the perturbed potential, and $E_n = \varepsilon_n + \Delta \varepsilon_n$. Then, Eq. (5), and subsequently Eq. (3), reduces to

$$(W_n + \Delta W_n)^2 - (W_n + \Delta W_n)' = V - E_n \ , \tag{9}$$

which is similar to Eq. (1), nevertheless it is valid for all states unlike usual supersymmetric treatments [29] which use (9) only for the ground state due to theoretical considerations. Further, as one in principle knows explicitly the solution of Eq. (7), namely the whole spectrum and corresponding eigenfunctions of the unperturbed interaction potential, the goal here is to solve only Eq. (8), which is the backbone of this formalism. The reader is referred to [11] for the successful applications of (8) involving different problems in quantum theory through exactly solvable potentials.

However, if the whole potential has no analytical solution as the case considered in this Letter, which means $\Delta W$ and subsequently Eq. (8) cannot be exactly solvable, then one can expand the functions in terms of the perturbation parameter $\lambda$ ,

$$\Delta V(r;\lambda) = \sum_{N=1}^{\infty} \lambda^N \Delta V_N(r), \quad \Delta W_n(r;\lambda) = \sum_{N=1}^{\infty} \lambda^N \Delta W_{nN}(r), \quad \Delta \varepsilon_n(\lambda) = \sum_{N=1}^{\infty} \lambda^N \varepsilon_{nN} \qquad (10)$$

where $N$ denotes the perturbation order. Substitution of the above expansion into Eq. (8) by equating terms with the same power of $\lambda$ on both sides yields up to for instance $O(\lambda^3)$

$$2W_n \Delta W_{n1} - \Delta W'_{n1} = \Delta V_1 - \Delta \varepsilon_{n1} \quad , \qquad (11)$$

$$\Delta W_{n1}^2 + 2W_n \Delta W_{n2} - \Delta W'_{n2} = \Delta V_2 - \Delta \varepsilon_{n2} \quad , \qquad (12)$$

$$2(W_n \Delta W_{n3} + \Delta W_{n1} \Delta W_{n2}) - \Delta W'_{n3} = \Delta V_3 - \Delta \varepsilon_{n3} \quad , \qquad (13)$$

Eq. (8) and its expansion give a flexibility for the easy calculations of the perturbative corrections to energy and wave functions for the *nth* state of interest through an appropriately chosen perturbed superpotential. It has been shown [11] that this feature of the present model leads to a simple framework in obtaining the corrections to all states without using complicated mathematical procedures.

## 3. APPLICATION

For clarity, in this paper we restrict ourselves to the Schrödinger equation in one dimension $(\ell = 0)$ and consider the anharmonic potential as

$$V = V_0 + \Delta V = x^2 + gx^4 \quad , \qquad (14)$$

in which the unperturbed potential represents the well known factorizable harmonic oscillator. From the literature [29,30], the corresponding superpotentials, wave functions and energy values are

$$W_n = -\sqrt{a} \left[ \sqrt{a} x - \frac{H_{n+1}(\sqrt{a} x)}{H_n(\sqrt{a} x)} \right] \quad , \quad \chi_n = H_n(\sqrt{a} x) \exp(-ax^2/2) \quad , \quad \varepsilon_n = 2a\left(n + \frac{1}{2}\right) \quad , \qquad (15)$$

where $H_n$ denotes the Hermite polynomials, $n = 0,1,2,...$ is the radial quantum number and $a$ is the potential parameter. With a suitable choice of $\Delta W$,

$$\Delta W = \sum_{N=1}^{\infty} f_N x^{2N+1} \quad , \qquad (16)$$

corresponding to the perturbed potential $gx^4$ in (14), one obtains some equations at successive orders for different states, which reveal some interesting relations between them leading to a simple algebraic treatment of the problem of interest here.

**3.1. Calculations for $n = 0$ and $n = 1$ states**

For instance, starting from the ground state calculations $(n = 0)$, where, from (15), $W = ax$ and considering Eqs. (11) through (13) we get at the first order $(N = 1)$,

$$2af_1 = g \quad , \quad f_1 = \frac{1}{3}(a^2 - 1) \quad \Rightarrow \quad E_{n=0}^3 - E_{n=0} - \frac{3}{2}g = 0 \ . \tag{17}$$

Similarly, at the second order $(N = 2)$ of the perturbation we have

$$f_1^2 + 2af_2 = 0 \quad , \quad f_2 = \frac{2af_1 - g}{5} \quad \Rightarrow \quad E_{n=0}^4 - \frac{22}{17}E_{n=0}^2 - \frac{18g}{17}E_{n=0} + \frac{5}{17} = 0 \ , \tag{18}$$

and the third order $(N = 3)$ calculations give

$$2(af_3 + f_1 f_2) = 0, \ f_3 = \frac{f_1^2 + 2af_2}{7} \quad \Rightarrow \quad E_{n=0}^5 - \frac{50}{31}E_{n=0}^3 - \frac{39g}{31}E_{n=0}^2 + \frac{19}{31}E_{n=0} + \frac{21g}{31} = 0 \tag{19}$$

If one repeats the same calculations for the first excited state $(n = 1)$, for which the superpotential is set $W = ax - \frac{1}{x}$ in the light of Eq. (15), then the first order yields

$$2af_1 = g \quad , \quad f_1 = \frac{1}{5}(a^2 - 1) \quad \Rightarrow \quad E_{n=1}^3 - 9E_{n=1} - \frac{135}{2}g = 0 \ , \tag{20}$$

and at the second order we have

$$f_1^2 + 2af_2 = 0 \quad , \quad f_2 = \frac{2af_1 - g}{7} \quad \Rightarrow \quad E_{n=1}^4 - \frac{34}{3}E_{n=1}^2 - 50gE_{n=1} + 21 = 0 \ , \tag{21}$$

while the third order expressions are

$$2(af_3 + f_1 f_2) = 0, \ f_3 = \frac{f_1^2 + 2af_2}{9} \quad \Rightarrow \quad E_{n=1}^5 - 14E_{n=1}^3 - 57gE_{n=1}^2 + 45E_{n=0} + 243g = 0 \ . \tag{22}$$

In our calculations, the upper bounds which are the largest real and positive roots in these equations are chosen as the energy of the anharmonic oscillator in the related quantum state.

The repeat of such calculations for large succesive orders reproduces similar relations in a manner of hieararchy. The systematic calculation of perturbation corrections of large orders offer no difficulty if we resort a computer algebra system like Mathematica, Mapple or Reduce. This realization leads us to generalize anharmonic oscillator solutions for the ground and first excited states without solving the Schrödinger equation. To calculate the energy values individually at each perturbation order, one needs to solve only

$$\sum_{k=0}^{N} f_k f_{N-k} - g\delta_{N1} = 0 \ , \tag{23}$$

in which δ is the Kronecker delta and $f_0 = a$ is the parameter related to Eq. (15). The perturbation coeffients above can easily be computed through

$$f_N = (2N + 2n + 1)^{-1} \left( \sum_{k=0}^{N-1} f_k f_{N-k-1} - \delta_{N1} - g\delta_{N2} \right) . \qquad (24)$$

The calculations are carried out for different range of $g$ values and the results obtained for the ground and first excited state energies are compared to the one computed numerically [19]. The agreement is remarkable in the whole range of $g$ values for the both quantum state, see Tables 1 and 2. The large order perturbation calculations are performed by a simple use of Mathematica [31] along the line of (23) and (24) with simple algebraic manipulations.

### 3.2. Calculations for $n \geq 2$ states

When dealing with excited states this approach seems rather cumbersome because the zeros of the wavefunction have to be taken into account explicitly. However, with some simple but physically acceptable algebraic manipulations, we can obtain simple analytical expressions for higher excited states easily from a straightforward generalisation of the resulting expressions at successive perturbation orders as in the previos section.

Starting with the second excited state $(n = 2)$, where from (15) the superpotential is $W_{n=2} = ax(2ax^2 - 5)/(2ax^2 - 1)$, energies up to for example the fifth order $(N = 5)$ can be obtained through

$$2af_1 = g \quad , \quad f_1 = \frac{1}{8}(a^2 - 1) \quad , \quad N = 1 \quad ,$$

$$f_1^2 + 2af_2 = 0 \quad , \quad f_2 = \frac{2af_1 - g}{10} \quad , \quad N = 2 \quad ,$$

$$2(af_3 + f_1 f_2) = 0 \quad , \quad f_3 = \frac{f_1^2 + 2af_2}{12} \quad , \quad N = 3 \quad ,$$

$$f_2^2 + 2(f_1 f_3 + af_4) = 0 \quad , \quad f_4 = \frac{2(af_3 + f_1 f_2)}{14} \quad , \quad N = 4 \quad ,$$

$$2(f_2 f_3 + f_1 f_4 + af_5) = 0 \quad , \quad f_5 = \frac{f_2^2 + 2(f_1 f_3 + af_4)}{16} \quad , \quad N = 5 \quad . \qquad (25)$$

In these treatments, to remove the singularities in the related superpotential due to the zeros of the wavefunction, we accept that $2ax^2 \succ 1$ leading to physically acceptable results. This simple assumption reproduces good accuracy in the calculations when compared to tedios calculations of LPT for higher excited states. The results obtained are shown in Table 3.

Finally, within the same framework one can readily get similar expressions for other excited states. For instance, in order to deal with the third excited state $(n=3)$ of the anharmonic oscillator one can choose the corresponding superpotential $W_{n=3} = (2a^2 x^4 - 9ax^2 + 3)/(2ax^3 - 3x)$ via (15) for the unperturbed piece of the potential and end up with some explicit algebraic equations at each order,

$$2af_1 = g \quad , \quad f_1 = \frac{1}{12}(a^2 - 1) \quad , \quad N = 1 ,$$

$$f_1^2 + 2af_2 = 0 \quad , \quad f_2 = \frac{2af_1 - g}{14} \quad , \quad N = 2 ,$$

$$2(af_3 + f_1 f_2) = 0 \quad , \quad f_3 = \frac{f_1^2 + 2af_2}{16} \quad , \quad N = 3 , \tag{26}$$

…and so on. In this case the wave function and consequently the superpotential have three zeros at $x = \pm\sqrt{3/2a}$. As argued above, to circumvent the resulting singularities the present calculations here make a similar assumption that is $2ax^3 \succ 3x$ which produces reasonable results. This choice however for higher excited states with $n \geq 3$ allows only the coefficients $f_N$ with $x^{2N}$ and $x^{2N+2}$ through the linear perturbation expressions at each order. The results obtained are illustrated in Table 3. Although the present formalism suggest a systematic way of improving the anharmonic oscillator perturbation series, the accuracy of the present formalae as expected gets decrease with the increase of the quantum number since the perturbation becomes more important. Nevertheless, owing to the nearly correct large–$g$ behaviour of the results presented here they are expected to be much more accurate than the perturbation series. This idea exploited by Fernandez et al [20] in order to obtain analytical expressions for the eigenvalues of the anharmonic oscillator from the semiclassical considerations.

In the light of the above discussions one can easily generalize the whole calculations discussed in section 3 in a compact form to determine the solutions of quartic anharmonic oscillator in a closed algebraic form, which should be valid for the all states. Eq. (23) can be safely used for this purpose, however the coefficients should be re-defined as

$$f_N = (2N + 2n + \alpha_n)^{-1} \left( \sum_{k=0}^{N-1} f_k f_{N-k-1} - \delta_{N1} - g\delta_{N2} \right) , \qquad (27)$$

where $\alpha_n = (n-1) + \alpha_{n-1}$ being with $n \geq 1$ and $\alpha_0 = 1$. As a matter of fact, the only data that is needed when using Mathematica is (27) to solve (23) yielding energy values through the perturbation orders for any quantum state.

### 3.3. Large-order calculations

A question now arises about the convergence of the method just described. Since it seems closely related to perturbation theory, one expects it to be asymptotic divergent. Our numerical results almost confirm this assumption. We have calculated low-lying energy levels of the anharmonic oscillator for several $g$ values, finding almost the same behaviour in all cases. Tables 5 and 6 represent the oscillations of our results, though they remain quite close to the true eigenvalue, about its actual value as the perturbation order $(N)$ increases, which are carried out for $g = 1$ and $g = 10$ respectively for the lowest state. Although divergent the present method is still useful because it certainly improves the perturbation series. The most accurate results is obtained from the $N$ value corresponding to the smallest oscillation amplitude. Such an accuracy cannot be obtained from the other perturbation series.

### 4. CONCLUDING REMARKS

We have shown that the eigenvalues of quantum mechanical systems can be approximately obtained from the present formalism which is non-perturbative, self-consistent and systematically improvable. Although we have limited ourselves to one illustrative example, the range of application of the method is rather large and appears to be straightforward. The perturbation procedure is well adapted to the use of software systems such as Mathematica and allows the computation to be carried out up to high orders of the perturbation. For any given state, simple algebraic manipulations provide, at the same time, analytical expressions of the perturbed eigenvalues and eigenfunctions, without having to compute any matrix elements or to perform any integration.

The increase in the value of $g$ for different quantum numbers do not imply special difficulty since the perturbed contributions merely follow from the solution of a linear system of equations of small order. Within this context, we may for example recall that Hill determinants of orders as high as $100 \times 100$ are required [18] for large values of $g$ $(g \approx 50)$ and that, when applying summation produres, the calculations becomes more and more cumbersome as $g$ increases, because of the strong divergence of the coefficients in the Rayleigh Schrödinger expansion. Furthermore, the remove of the singularities in the unperturbed wavefunction via the superpotential introduced in the present formalism does not cause tedious calculations which are great pain when dealing with excited states in LPT. Finally, although in this Letter we have focused only the calculations of eigenvalues for the quartic anharmonic oscillator, one can also find analytical solutions easily for the corresponding total wavefunction, if necessary, through the use of Eqs. (4), (6), (15) and (16).

As a concluding remark, due to its simplicity and accuracy in particular for small $g-$values at low-lying states we believe this method to be competitive with other methods developed to deal with perturbation treatments. As a matter of fact that, the degree of precision of the results can be drastically improved by raising the perturbative order in the expansion, a step which does not bear any technical difficulty. It would be interesting to extend the present scheme to other non-exactly solvable potentials.

**TABLES**

**Table 1.** Lowest eigenvalue of the anharmonic oscillator $(n = 0)$

| $g$ | N=1 | N=2 | N=3 | N=4 | Exact [19] |
|---|---|---|---|---|---|
| 0.001 | 1.00075 | 1.00075 | 1.00075 | 1.00075 | 1.000748 |
| 0.01 | 1.00742 | 1.00737 | 1.00737 | 1.00737 | 1.007373 |
| 0.05 | 1.03558 | 1.03467 | 1.03474 | 1.03473 | 1.034729 |
| 0.1 | 1.06792 | 1.06500 | 1.06533 | 1.06528 | 1.065286 |
| 0.5 | 1.26255 | 1.23689 | 1.24347 | 1.24118 | 1.2418541 |
| 1.0 | 1.43113 | 1.38082 | 1.39672 | 1.39017 | 1.392352 |
| 10 | 2.60124 | 2.38404 | 2.47867 | 2.42910 | 2.449174 |
| 100 | 5.37603 | 4.82115 | 5.08211 | 4.93770 | 4.999417 |
| 1000 | 11.4763 | 10.2346 | 10.8285 | 10.4960 | 10.639789 |
| 10000 | 24.6756 | 21.9784 | 23.2731 | 22.5463 | 22.861608 |

**Table 2.** First excited state energies of the anharmonic oscillator $(n = 1)$

| $g$ | N=1 | N=2 | N=3 | N=4 | N=8 | Exact [19] |
|---|---|---|---|---|---|---|
| 0.001 | 3.00374 | 3.00374 | 3.00374 | 3.00374 | 3.00374 | 3.003739 |
| 0.01 | 3.03682 | 3.03652 | 3.03653 | 3.03653 | 3.03653 | 3.036525 |
| 0.05 | 3.17236 | 3.16683 | 3.16727 | 3.16722 | 3.16723 | 3.167225 |
| 0.1 | 3.32148 | 3.30511 | 3.30718 | 3.30681 | 3.30687 | 3.306872 |
| 0.5 | 4.14123 | 4.03032 | 4.05869 | 4.04924 | 4.05171 | 4.051932 |
| 1.0 | 4.80180 | 4.60453 | 4.66448 | 4.64159 | 4.64784 | 4.648813 |
| 10 | 9.11388 | 8.39998 | 8.68054 | 8.55128 | 8.58582 | 8.599004 |
| 100 | 19.0576 | 17.3193 | 18.0446 | 17.6965 | 17.7864 | 17.83019 |
| 1000 | 40.7899 | 36.9427 | 38.5693 | 37.7818 | 37.9829 | 38.08683 |
| 10000 | 87.7547 | 79.4176 | 82.9526 | 81.2378 | 81.6747 | 81.90331 |

**Table 3.** Second excited state energies of the anharmonic oscillator $(n = 2)$

| $g$ | N=1 | N=2 | N=3 | N=4 | N=15 | Exact [19] |
|---|---|---|---|---|---|---|
| 0.001 | 5.00997 | 5.00996 | 5.00996 | 5.00996 | 5.00996 | 5.009711 |
| 0.01 | 5.09715 | 5.09606 | 5.09609 | 5.09609 | 5.09609 | 5.093939 |
| 0.05 | 5.44017 | 5.42257 | 5.42423 | 5.42401 | 5.42404 | 5.417261 |
| 0.1 | 5.79852 | 5.75129 | 5.75799 | 5.75670 | 5.75694 | 5.747959 |
| 0.5 | 7.60690 | 7.35517 | 7.41992 | 7.39911 | 7.40489 | 7.396900 |
| 1.0 | 8.98161 | 8.56694 | 8.68960 | 8.64563 | 8.65908 | 8.655049 |
| 10 | 17.5870 | 16.2662 | 16.7452 | 16.5461 | 16.6188 | 16.63592 |
| 100 | 37.0665 | 33.9532 | 35.1363 | 34.6287 | 34.8238 | 34.87398 |
| 1000 | 79.4750 | 72.6342 | 75.2605 | 74.1261 | 74.5674 | 74.68140 |
| 10000 | 171.046 | 156.245 | 161.940 | 160.830 | 160.437 | 160.6859 |

**Table 4.** Third excited state energies of the anharmonic oscillator $(n = 3)$

| $g$ | N=1 | N=2 | N=3 | N=4 | N=15 | Exact [19] |
|---|---|---|---|---|---|---|
| 0.001 | 7.02091 | 7.02087 | 7.02087 | 7.02087 | 7.02087 | 7.018652 |
| 0.01 | 7.20124 | 7.19823 | 7.19833 | 7.19832 | 7.19832 | 7.178573 |
| 0.05 | 7.87793 | 7.83590 | 7.84053 | 7.83985 | 7.83995 | 7.770271 |
| 0.1 | 8.54838 | 8.44564 | 8.46179 | 8.45849 | 8.45913 | 8.352677 |
| 0.5 | 11.7019 | 11.2511 | 11.3683 | 11.3315 | 11.3415 | 11.11515 |
| 1.0 | 14.0000 | 13.2973 | 13.5021 | 13.4319 | 13.4524 | 13.15680 |
| 10 | 28.0000 | 25.9479 | 26.6524 | 26.3804 | 26.4698 | 25.80627 |
| 100 | 59.3169 | 54.5806 | 56.2681 | 55.5997 | 55.4001 | 54.38529 |
| 1000 | 127.327 | 116.968 | 120.689 | 119.207 | 119.712 | 116.60319 |
| 10000 | 274.100 | 251.711 | 259.767 | 256.555 | 257.651 | 250.95073 |

**Table 5.** Lowest eigenvalues calculated for $g = 1$ at large orders

| $N$ | $E_{n=0}$ | $N$ | $E_{n=0}$ |
|---|---|---|---|
| 5 | 1.39357 | 15 | 1.39269 |
| 6 | 1.39155 | 16 | 1.39196 |
| 7 | 1.39291 | 17 | 1.39272 |
| 8 | 1.39191 | 18 | 1.39221 |
| 9 | 1.39271 | 19 | 1.39273 |
| 10 | 1.39202 | 20 | 1.39231 |
| 11 | 1.39265 | 21 | 1.39273 |
| 12 | 1.39201 | 22 | 1.39235 |
| 13 | 1.39266 | 23 | 1.39272 |
| 14 | 1.39186 | 24 | 1.39238 |
| $E_{n=0}^{exact} = 1.392352$ | | | |

**Table 6.** Lowest eigenvalues calculated for $g = 10$ at large orders

| $N$ | $E_{n=0}$ | $N$ | $E_{n=0}$ |
|---|---|---|---|
| 5 | 2.46214 | 15 | 2.45815 |
| 6 | 2.43752 | 16 | 2.44941 |
| 7 | 2.45804 | 17 | 2.45808 |
| 8 | 2.43856 | 18 | 2.45067 |
| 9 | 2.45720 | 19 | 2.45800 |
| 10 | 2.43125 | 20 | 2.45176 |
| 11 | 2.45752 | 21 | 2.45798 |
| 12 | 2.44277 | 22 | 2.45276 |
| 13 | 2.45799 | 23 | 2.45798 |
| 14 | 2.44735 | 24 | 2.45358 |
| $E_{n=0}^{exact} = 2.449174$ | | | |